\documentclass[]{spie}  
\usepackage[dvips]{graphicx}

\title{Active Optics on the Baade 6.5-m (Magellan I) Telescope} 

\author{Paul L. Schechter\supit{a}, 
Greg Burley\supit{b},
Charles L. Hull\supit{b},
Matt Johns\supit{b},
Buddy Martin\supit{c},\\
Skip Schaller\supit{d},
Stephen A. Shectman\supit{b},
and Steven C. West\supit{e}
\skiplinehalf
\supit{a}Massachusetts Institute of Technology, Cambridge,
MA 02139-4307, USA \\
\supit{b}Carnegie Observatories, Pasadena CA, 91101-1209 USA \\
\supit{c}Steward Observatory, University of Arizona, Tucson AZ 85721 USA \\
\supit{d}Las Campanas Observatory, Casilla 601, La Serena, Chile \\
\supit{e}MMT Observatory, 933 N. Cherry Avenue, Tucson AZ 85719  \\
}

\begin{document} 
\newcommand\farcs{\mbox{$.\!\!^{\prime\prime}$}}%
  \maketitle 
\begin{abstract}
The Magellan active optics system has been
operating continuously on the Baade 6.5-m since the
start of science operations in February 2001.  The
active optical elements include the primary
mirror, with 104 actuators, and the secondary
mirror, with 5 positional degrees of freedom.
Shack-Hartmann (SH) wavefront sensors are an
integral part of the dual probe guiders.
The probes function interchangeably, with either
probe capable of guiding or wavefront sensing.  In
the course of most routine observing stars
brighter than 17th magnitude are used to apply
corrections once or twice per minute.  The rms
radius determined from roughly 250 SH spots
typically ranges between 0\farcs05 and 0\farcs10.
The spot pattern is analyzed in terms of a mixture
of 3 Zernike polynomials (used to correct the
secondary focus and decollimation) and 12 bending
modes of the primary mirror (used to compensate
for residual thermal and gravitational
distortions).  Zernike focus and the lowest order
circularly symmetric bending mode, known
affectionately as the ``conemode," are sufficiently
non-degenerate that they can be solved for and
corrected separately.
\end{abstract}



\section{INTRODUCTION}

Almost every major optical telescope of the current generation has an
actively controlled primary mirror support system.  No matter how much
thought and care may have gone into the design of a telescope, the
ultimate test is the quality of the images it delivers to the focal
plane.  Wavefront sensors can both monitor a mirror support system and
provide feedback as needed.  An excellent review of the use of such
systems is given by Noethe\cite{Noethe01}.

The Magellan I (Baade) and II (Clay) telescopes have f/1.25 primaries,
making the tolerances on the secondary position exceedingly tight.
Furthermore the figures of the 6.5-m borosilicate primary mirrors are
sensitive to temperature changes.  Some 4 years before first light
the decision was taken to incorporate wavefront sensors into the
facility guiders, to ensure that focus, collimation and the figure of
the primary mirror are maintained through the course of routine
operations.  At the Baade telescope it is standard procedure to
acquire both a traditional guide star and a wavefront star for every
observation.

While the Magellan active optics system is quite simple, it appears to
work effectively and reliably.  It may serve as an example of how
little one can do and still produce good images.  The primary mirror
support system is described in \S2.  The guider is described in \S3.
Wavefront analysis is described in \S4.  Operation and performance are
described in \S5.

\section{MIRROR SUPPORT SYSTEM}

The primary mirror support for the Magellan telescopes is similar to
that for the MMT, which is described in detail by Martin {\it et
al.}\cite{MartCC98}  What follows is drawn from their more extensive
discussion.  Axial support is supplied by 104 pairs of opposed
push-pull air cylinders, each with a load of $\sim 1000$ N at zenith,
under the control of a Motorola VME 68K computer running VXWORKS.
These can be adjusted in $1$ N increments.  Feedback is supplied by
load cells located on three hard points, with an outer control loop
nulling the forces on these.  Changes due to gravity, telescope
accelerations and wind loading are corrected with a bandwidth of 1 Hz.

A finite element analysis of the primary was carried out by BCV
Progetti, showing how a force of $100$ N applied to each of the
actuators influenced 3200 locations on the surface.  The finite
element analysis was verified in a series of tests at the Steward
Observatory Mirror Lab, showing that the surface responds as expected
to the forces applied.  A singular value decomposition (SVD) was
carried out to identify a sequence of ``bending modes'' ordered from
softest to stiffest.  Each bending mode consists of a set of forces
and the correspoding displacement of the mirror.  Both the force
vectors and the surface displacements are orthogonal.\footnote {Our
bending modes are similar to the minimum elastic energy modes
discussed by Noethe\cite{Noethe01}.  They differ in that our SVD modes
minimize forces on a discrete set of actuators rather than the elastic
energy of the mirror.}

The surface distortions of the low order modes correspond quite
closely to Zernike polynomials, with the exception of the third
softest (lowest order axisymmetric) mode.  Martin {\it et al.} note
that the surface deflection for this mode is more nearly conical than
the paraboloidal shape expected for Zernike focus, and that it
includes a substantial element of Zernike spherical aberration.

An early study showed that the forces required to produce low order
Zernike polynomial displacements on the surface of the mirror could be
very much reduced by restricting the force set to the 32 softest
mirror modes, with very little loss in surface accuracy.  During the
first half year of operation, when we analyzed the wavefront in terms
of Zernike polynomials, we projected the wavefront onto these low
order modes and ignored the projections onto the stiffer modes in
computing correction forces.  We subsequently analyzed the wavefront
directly in terms of the bending modes, using a yet smaller subset.
\begin{figure}
\begin{center}
\vspace{6.0truein}
\includegraphics{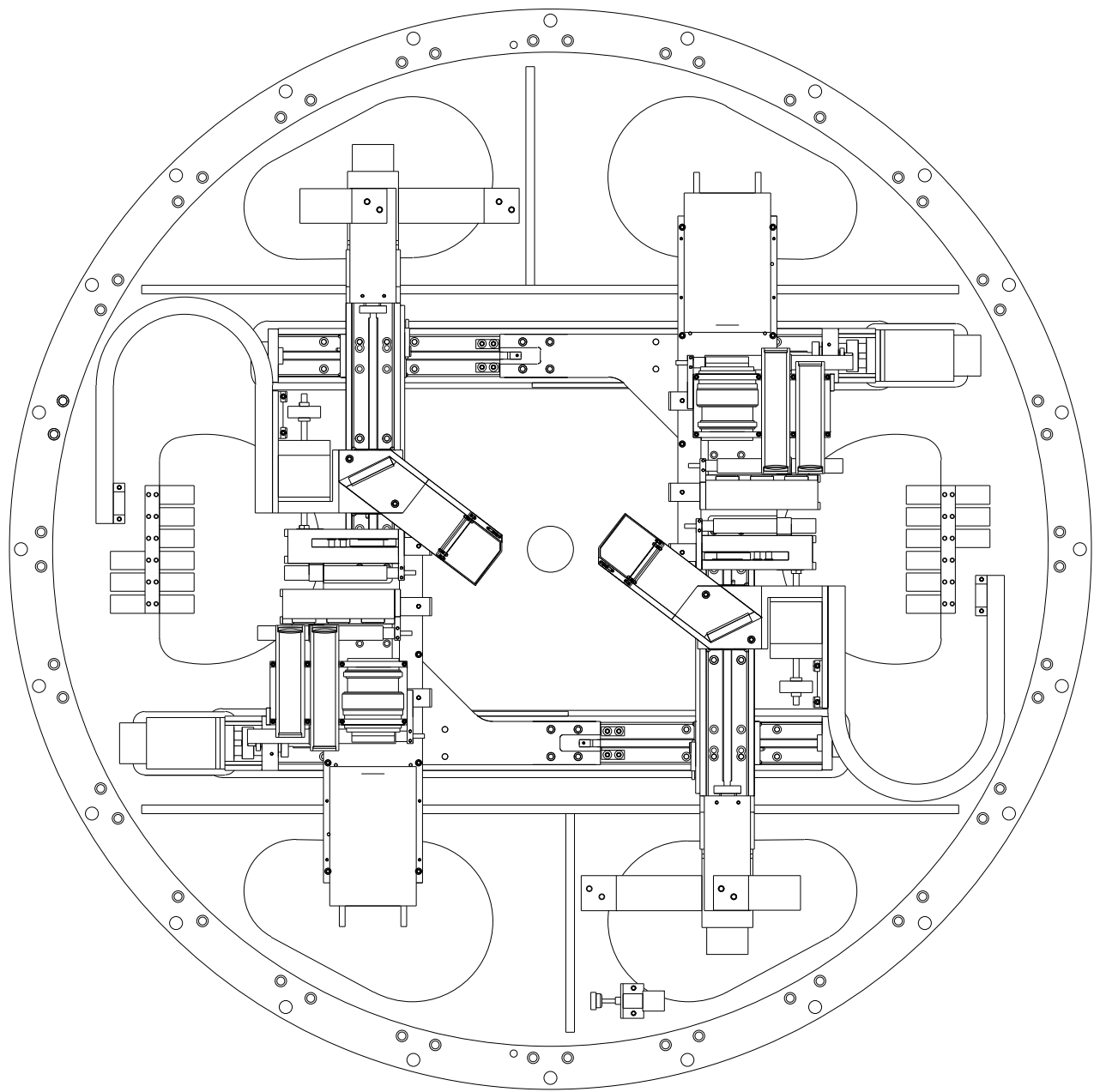}
\end{center}
\caption[example] 
{\label{fig:guider} The Magellan guider.  The pickoff mirrors send the
beam up and to the left (or down and to the right) and then back down
(up) to the optics assemblies.  The rectangles at 1 and 7 o'clock are
the two cameras.  In front of them are the 2.5:1 reduction lenses used
for direct guiding.  Two cylindrical Shack-Hartmann assemblies lie next to
each lens.}
\end{figure} 

\section{FACILITY GUIDERS}

Two facility guiders are currently in operation at the f/11 Nasmyth
focii of the Baade telescope, with two more to be installed on the
Clay telescope by the end of 2002.  Each telescope has three
additional folded Gregorian ports.  Eventually a total of six guiders
will be mounted on the two telescopes for use with small and
medium-sized instruments.  Large instruments will have their own
internal guider optics and wavefront sensors but will make use of the
hardware and control software developed for the facility guiders.

The guiders, as shown in Figure \ref{fig:guider}, have two identical probe
assemblies, symmetrically arranged, each consisting of a pickoff
mirror, three 3-position cross-slides that carry optics, and a CCD
camera. The cross-slides are moved by pneumatic pistons.  The first
cross-slide in the optical train is located at the conjugate telescope
focus and lets the operator select either a field lens for direct
guiding, an aperture for use with the wavefront analyzers, or a
calibrator for the wavefront analyzer.  The calibrator consists of a
100 micron pinhole and an LED illuminator.  A filter slide with red
and blue filters and an open position comes next in the optical train.
The third slide carries a re-imaging lens for direct guiding and a
choice of two wavefront analyzers for coarse and fine spatial sampling
of the telescope pupil.  The re-imaging lens is an Olympus Zuiko 50 mm
macro lens that gives a 2.5:1 reduction at the CCD.

Each Shack-Hartmann wavefront analyzer consists of a 100 mm focal
length collimating lens and a lenslet array.  The collimating lens
forms a 9 mm diameter image of the telescope pupil on the array.
Currently the choice of arrays gives approximately 12 and 18 samples
across the pupil for the ``coarse" and ``fine" analyzers respectively
although in practice only the fine array is used.  It has a pitch of
0.5 mm, a 30 mm focal length and a $20 \times 20$ format that allows some
motion of the pupil image as the probe moves around the field before
the pupil falls off the array.

The probe assemblies are mounted on perpendicular x-y recirculating
ball slides that allow the pick-off mirrors to move around the field
of view.  Ball screws with stepper motors and encoders control the
slide positions.  Each probe can cover a region 7.7 arcmin by 12.9
arcmin on its side of the guider.  The regions overlap so that both
probes have access to the center of the field.  A focus mechanism with
25 mm travel between the pick-off mirror and the first cross-slide
compensates for focal plane curvature.  The instrument focus behind
the guider assembly is nominally fixed at 125 mm.

Each CCD camera can be used for both direct guiding and wavefront
analysis, depending upon the configuration of the probe cross-slides.
Each camera has a thermoeletrically cooled, back-illuminated Marconi
$1024 \times 1024$ frame-store CCD with 13 $\mu$m pixels.  These are
always binned $2
\times 2$.  A total of 20 such cameras have been constructed for use
in the facility guiders and in intrument specific applications.  The
guide cameras are controlled by PCs running DOS.

\begin{figure}
\begin{center}
\vspace{4.70 truein}
\includegraphics{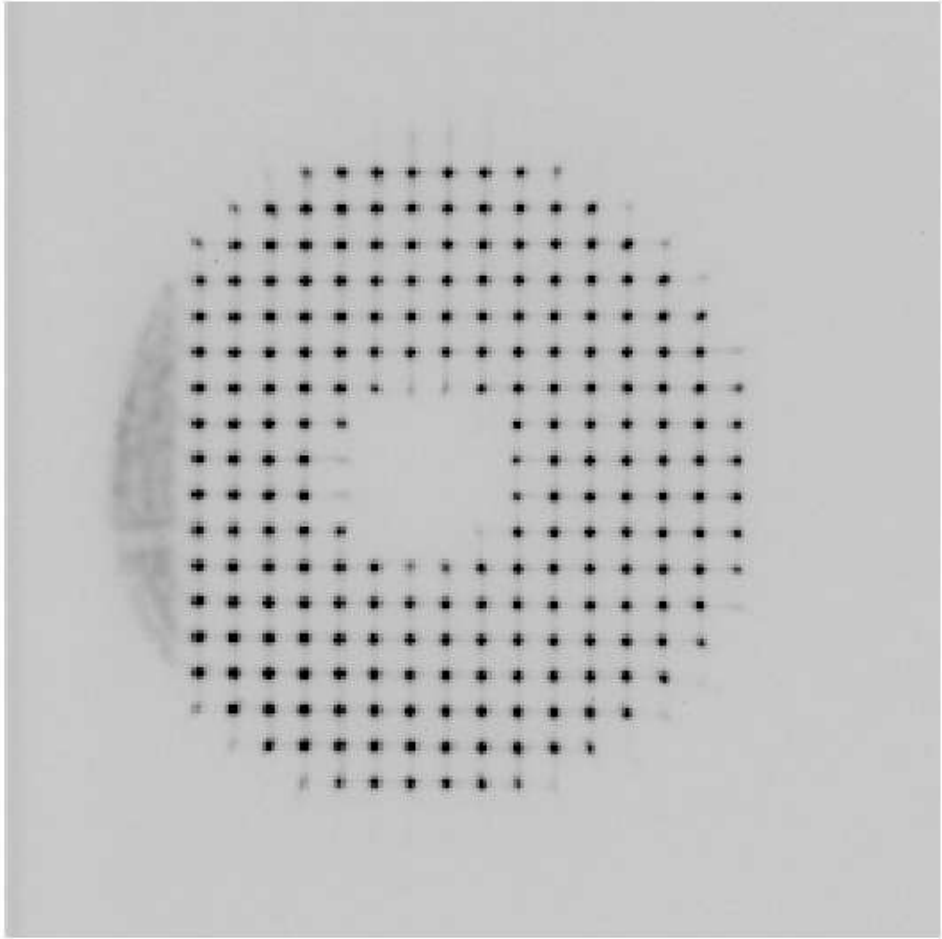}
\end{center}
\caption[example] 
{\label{fig:spots} A Shack-Hartmann image.  The spacing between spots
is $\sim$ 19 pixels.  To the left the pupil has run off the edge of
the lenslet array.  Partially illuminated lenslets at the outer and
inner edges of the pupil produce spots that are somewhat fainter, and
in some cases distorted.}
\end{figure}

In Figure \ref{fig:spots} we show a typical Shack-Hartmann image.  As the
probe is fairly close to the edge of the field, a portion of the pupil
falls off the lenslet array.

\section{WAVEFRONT ANALYSIS}
\subsection{Wavefront Errors}

For stars with $m \le 16$ mag, we use a combination of on-chip
integration and leaky-memory averaging to generate an effective
integration time of about 30 s on Shack-Hartmann exposures.  These
are compared with a stored exposure of the calibration LED.  Hot
pixels are flagged but the data are not otherwise corrected for
photometric sensitivity.  A customized version of the stellar
photometry program {\sc DoPHOT} \cite{SchecMS93} is used to find the
centroids and central intensities of all spots.  In a first pass
through the data a typical shape is determined for the point spread
function (PSF).  Spots that deviate significantly from this shape are
excluded from further consideration.  On a second pass through the
data the typical PSF is then used for all spots.  By fitting a model
PSF rather than computing centroids, accurate spot positions can be
obtained even when the data is corrupted by bad pixels.

A feature of our design is that the image of the pupil on the lenslet
array moves off the center as the pickoff mirror moves away from
the field center.  The pupil center is determined from intensity
measurements of those spots at the edge and center of the pupil formed
by lenslets that are only partially illumnated.  A model is fit to
these to obtain the pupil radius and center, with the latter accurate
to roughly 1 pixel.

Vector spot displacements relative to the reference LED exposure are
measured in pixels.  The positions of these spots, normalized by the
pupil radius, are computed with respect to the pupil center.  A set of
basis functions, {\it e.g.} Zernike polynomials, is used to
describe the wavefront.  These are differentiated and the resulting
slopes are used to obtain a best fit to the vector spot displacements
(typically some $\sim$ 220 of them) using a linear least squares
algorithm.\footnote{While Zernike polynomials are orthogonal, their
gradients are not.  The same holds true of the bending modes, as they
are constructed from surface displacements rather than slopes.
Nonetheless the covariances are small enough that the coefficients are
well determined.}  The coefficients of the basis functions can be
converted to wavefront errors using the parameters of the optical
system.\cite{Noethe01}

Without the wide-field corrector in the telescope, stars well off-axis
show strong Zernike coma and astigmatism.  We correct for these by
subtracting predicted vector spot displacements from the observed
displacements prior to fitting.

All of the above is carried out on a PC running Linux.  The same
machine serves as the telescope operator's interface to the telescope
control computer, a PC running DOS.  

\subsection{Correction}

At least some of the errors in the wavefront arise from the
displacement of the secondary mirror rather than the distortions of
the primary mirror.  In particular, we attribute the Zernike focus and
coma terms to the secondary and adjust it accordingly.  Ordinarily no
correction is made for wavefront tilt (pointing), as this is usually
under the control of the other guide probe.  The remaining terms, 12
at present, are used to construct a surface error attributed to the
primary mirror.  Force adjustments are computed for the primary mirror
actuators, and if the accumulated forces do not exceed a specified
limit -- typically 15\% of the nominal force -- the corrections are
applied.  If the force limit is exceeded, operator intervention is
required.

The time to transfer a Shack-Hartmann exposure and compute corrections
is roughly 5 s, during which time the next SH exposure is
accumulating.  Moreover the guide camera runs in a leaky average mode,
with a time constant equal to the exposure time.  We therefore set the
steady state gain, the applied fraction of the computed corrections at
33\%.  When an object is first acquired, after a slew, the errors can
be large.  To minimize setup time we set the gain to 100\% on the
first exposure, and reset the guide camera after applying these
corrections.

\subsection{Operator Interface}

The active optics software occupies one of four ``desktops'' on the
operator's monitor.  The operator sees the most recently acquired SH
image, with individual spot positions enclosed in green if a minimum
number of points has been identified (almost always the case), and in
red otherwise.  A vector plot of spot displacements is also displayed.
A spot diagram, computed from the displacement vectors, shows the size
of the image the telescope would have produced in the absence of
seeing, windshake, guiding errors and diffraction.  The accumulated
forces on the actuators are displayed graphically, as are the
amplitudes of the accumulated corrections in each basis function.

A text window reports on the progress of each frame.  The maximum
force (relative to the nominal actuator forces) and rms spot size are
also reported, as are the motions requested of the secondary.  The
amplitude of each basis function is given in terms of rms microns of
wavefront.  Typical values for astigmatism and focus might be 25 nm
in each mode, with values decreasing into the small single digits for
the highest modes fit.

\subsection{Choice of Basis Functions}

In our initial deployment we chose to expand the wavefront in terms of
Zernike polynomials, as these were easy to compute and readily
interpreted.  Over the course of the first few months of science
operations we found that our self-imposed force limits were frequently
exceeded, with the largest forces due to accumulated spherical
aberration.  This typically occurred at the beginning of the night,
and by midnight the spherical terms would usually disappear.  Our
attempts to make sense of this were inconclusive but we suspected that
we might do better using the bending modes as basis functions
rather than Zernike polynomials, which can be quite steep at the edge
of the pupil.

Though the bending modes might be better for the primary mirror,
Zernikes (focus and coma) are more appropriate for the secondary.  We
therefore switched to a hybrid scheme.  At low order the Zernike
polynomials and mirror modes are easily matched.  They have similar
topographies, with the same azimuthal structure and similar radial
structure.  It seemed likely that including both would cause the
solution to be unstable.  We adopted a flexible system in which one
might fit for one or the other or (at some risk) both.

Zernike coma and its associated mirror mode look so much like each
other that there was no hope of separating the two.  We fit for
Zernike coma and send corrections to the secondary.  But as discussed
in \S2, Zernike focus and its associated mirror mode are different.
Zernike focus has surface displacements that vary as the square of the
pupil radius, with slope errors proportional to pupil radius.  The
vector field of slope errors therefore looks like the Hubble expansion
of the universe.  By contrast the surface displacements of the
corresponding mirror mode vary very nearly linearly with radius.  The
slope errors, directed radially, all have the same length.

The fact that Zernike focus and its associated mirror mode can be
distinguished means that they can be solved for separately.  Of course
one might not want to fit for the corresponding mirror mode -- it
might be quite stiff and not easily generated by errors in the support
forces or thermal stresses.  But in our case this ``conemode'' is the
third softest, stiffer than the two astigmatic modes but softer than
trefoil.

\begin{table}
\caption{Bending modes and Zernike polynomials paired by symmetry.
The bending modes are identified by the numbers of azimuthal and
radial nodes, and are given in order of increasing stiffness.  The
Zernike modes are used to correct the translation of the secondary
while the bending modes are used to correct the primary.}
\label{basis}
\begin{center}       
\begin{tabular}{|c|cccccccc|}
\hline
bending &  2,0        & 0,0   &  3,0    &  1,1 & 4,0        & 2,1 & 5,0       &
           0,1      \\
\hline
Zernike & astigmatism & focus & trefoil & coma & quadrafoil & 5th
          order astig.  & pentafoil & spherical \\
\hline
\hline
used    &  2,0        & both  &  3,0    & coma & 4,0        & 2,1 & 5,0       &
           0,1       \\
\hline
\end{tabular}
\end{center}
\end{table}

This suggests a possible explanation for the large spherical
aberration we measured.  As noted by Martin {\it et
al.},\cite{MartCC98} the conemode projects onto all circularly
symmetric Zernikes, with the largest projection onto focus, a smaller
projection onto spherical and so forth.  Attempting to correct this
conemode by focusing the secondary leaves a significant amount of
uncorrected spherical and higher-order aberration.  While Zernike
focus may not be degenerate with the conemode taken alone, it will
become increasingly degenerate with some linear combination of
conemode and higher order circularly symmetric mirror modes as we add
more of the latter.  We have found empirically that our corrections
are stable if we include only the next higher mode, which corresponds
to Zernike spherical.

Reflecting the partial degeneracy between focus and conemode, the
corrections computed for both of these are correlated.  Suppressing
the corrections for conemode would reduce the computed focus
corrections by roughly a factor of two.

In Table 1 we show the mirror bending modes paired with the
corresponding Zernike polynomials of the same symmetry.  The modes are
ordered from softest to stiffest.  This ordering is similar to that of
the NTT\cite{Noethe01}, but with the 1,1 and 4,0 modes permuted.

Switching to mirror modes reduced the spot size by a factor of
$\sim$1.5, and the typical maximum force by roughly the same factor.
The force limits are exceeded less frequently.  When this does happen,
it is the 0,1 bending mode (corresponding to spherical aberration)
that produces the largest forces.  Fitting for this term is disabled
and typically re-enabled at the next pointing or a few hours later.

\section{OPERATION AND PERFORMANCE}

\begin{figure}[b]
\begin{center}
\vspace{3.25 truein}
\includegraphics{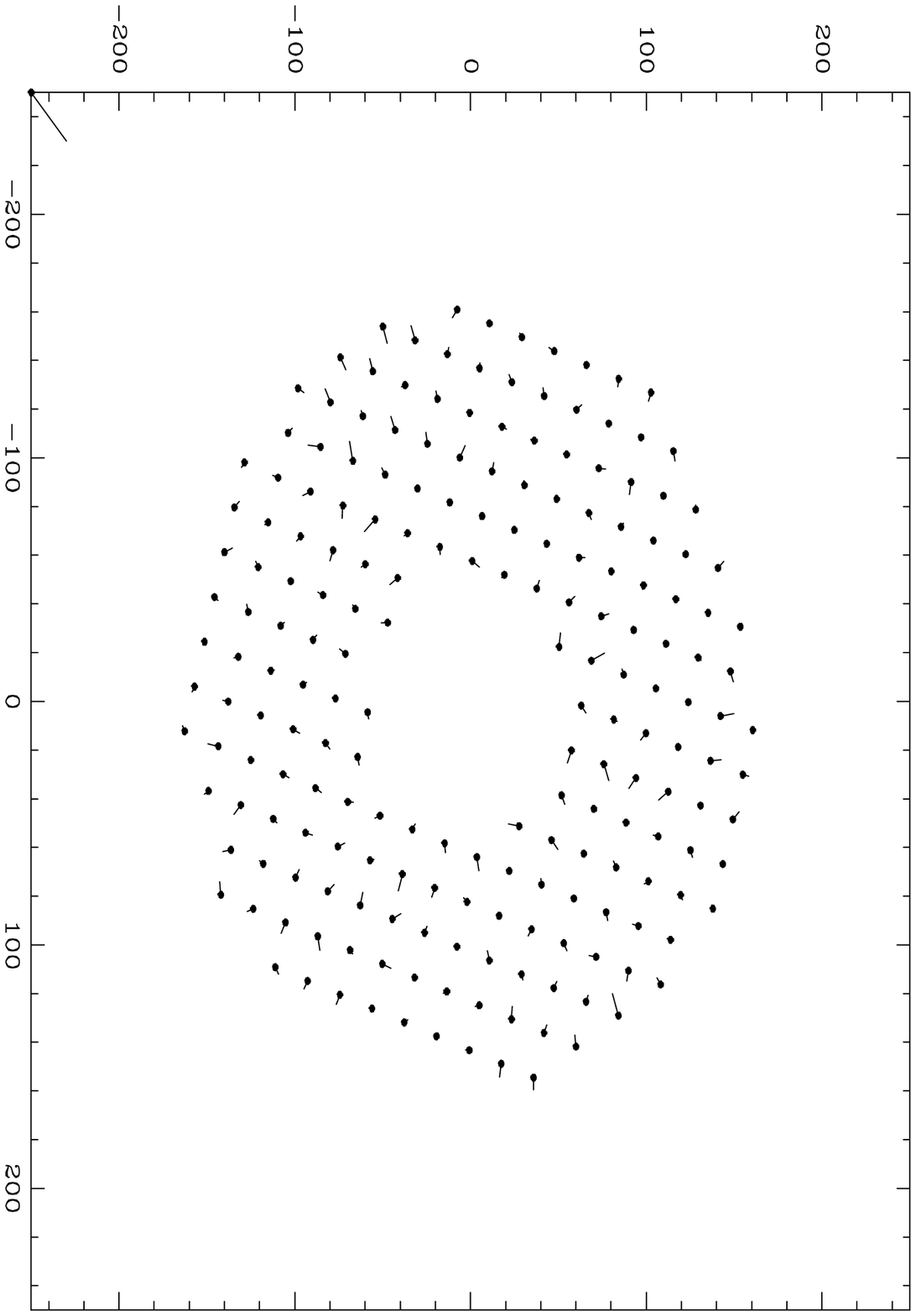}
\includegraphics{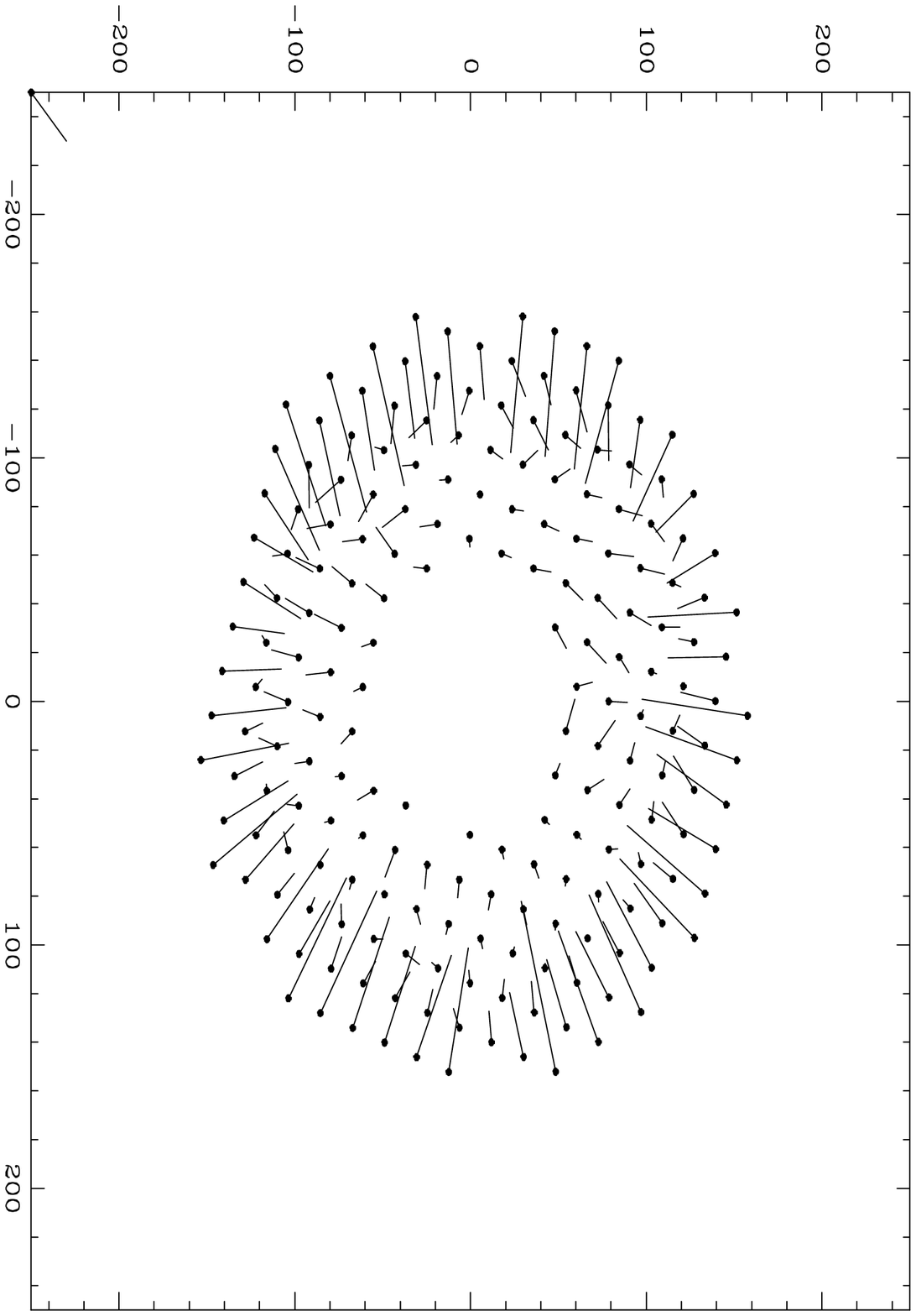}
\end{center}
\caption[example] 
{\label{fig:vectors} Spot displacements for two Shack-Hartmann images.
The displacements at the right were obtained on the first pointing
of the telescope at the beginning of a night.  Those at the left
were obtained at the end of the previous night.
}
\end{figure} 

Figure \ref{fig:vectors} shows spot displacement patterns for two
different SH images.  The one on the left was the last one obtained at
the end of a night.  The one at the right was the first one obtained
on the following night.  The tick mark to the lower left shows a
displacement of 0\farcs2.  The displacement field at the right is
not simply the product of defocus, which would produce displacements
proportional to distance from the center of the pupil.

These plots can be used to obtain a spot pattern.  The typical rms
spot pattern has an rms radius of 0\farcs075.  The best spot
patterns recorded have an rms of 0\farcs042.  This includes errors
in measuring spot positions.

The best optical exposures of 90s or longer obtained on the Baade
telescope have had a FWHM of 0\farcs30.  It would therefore seem
that the primary mirror and the position of the secondary contribute little
to the degradation of the delivered image quality, even in the best
conditions.

The active optics system operates sufficiently smoothly that some
observers are unaware of its existence.  Other, more experienced
observers initially react with skepticism when they are told that they
don't need to focus the telescope.  Probe offsets for each instrument,
putting the aperture of the SH analyzer at the focus of the telescope,
are measured and updated by the instrument specialists during
engineering runs.

\begin{figure}
\begin{center}
\vspace{3.25 truein}
\includegraphics{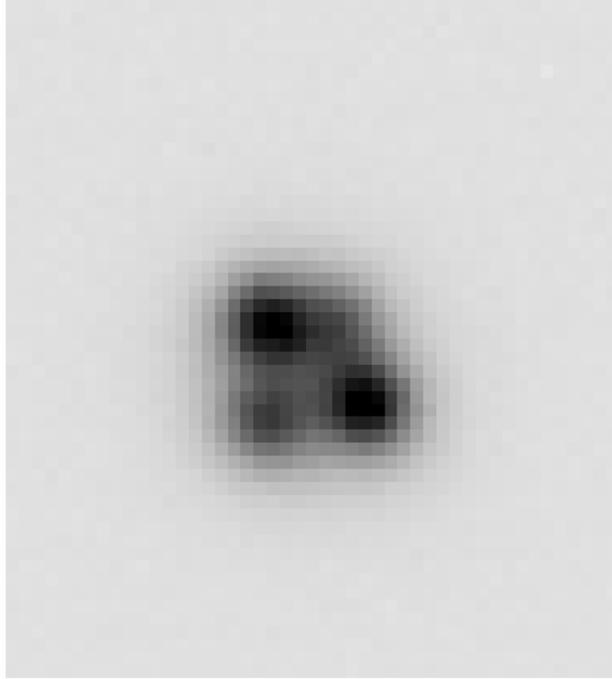}
\end{center}
\caption[example] 
{\label{fig:lens} A gravitationally lensed quasar discovered with the
Baade telescope.  The two brightest of the four quasar images are
separated by 0\farcs67.  The FWHM on this $1$ minute Sloan $i'$
exposure is 0\farcs41.}
\end{figure}

The end product is astronomical data.  In Figure \ref{fig:lens} we show a
$1$ minute exposure of a quasar drawn from the Hamburg-ESO digital
objective prism survey\cite{WisoCB00}.  It is one of three HE quasars
found to be gravitationally lensed in the first season of a survey with
Magellan.  An intervening galaxy produces a quartet of images, with
the two brightest separated by 0\farcs67.  The PSF on this image
has a FWHM of 0\farcs41.

\section{DISCUSSION}

While the implementation of the Magellan active optics system has gone
relatively smoothly, we learned some things in the process.
\begin{itemize}
\item{We found that we never use our coarse lenslet array.  Even at
high galactic latitude there is almost always a star bright enough to
correct the wavefront with the fine array.}
\item{While the pupil runs off the the lenslet array at large field
angles, the effect on correcting the wavefront seems minimal.}
\item{The use of bending modes rather than Zernike polynomials
reduced the forces required to correct the wavefront and gave better
correction.}
\item{With 18 Shack-Hartmann spots across the diameter of our pupil, we are
able to correct simultaneously for three azimuthally symmetric
wavefront terms: a focus term attributed to displacement of the
secondary; a ``conemode,'' the lowest circularly symmetric bending
mode of the primary; and the next softest bending mode corresponding
roughly to spherical aberration, but with admixtures of focus and higher
order modes.}
\end{itemize}

In the future we hope to incorporate larger lenslet arrays,
eliminating the current spillover.  Another possible improvement would
be use of SVD bending modes derived from mirror slopes rather than the
surface displacements.  These would give a set of basis functions that
is more nearly orthogonal for use in reconstructing the Shack-Hartmann
spot displacements.

\acknowledgments 
We thank Lothar Noethe for his counsel in the early stages of this
effort.  PLS gratefully acknowledges the support of the John Simon
Guggenheim Memorial Foundation and the hospitality of the Institute
for the Advanced Study.

\end{document}